\documentclass[showpacs,preprint,superscriptaddress,aps,pre]{revtex4}

\usepackage{amsmath}
\usepackage{amssymb}
\usepackage[T1]{fontenc}
\usepackage{graphicx}

\newcommand{\lds}{\tilde{\lambda}_\sigma}
\newcommand{\vRiz}{\mathbf{R}_i^0}
\newcommand{\vla}{\boldsymbol{\lambda}}
\newcommand{\vmu}{\boldsymbol{\mu}}
\newcommand{\vds}{{\boldsymbol{\delta}}^\sigma}
\newcommand{\vRis}{\mathbf{R}^\sigma_{{i}}}
\newcommand{\vRjt}{\mathbf{R}^\tau_{{j}}}
\newcommand{\vRtiaz}{\tilde{R}^0_{i,\alpha}}
\newcommand{\vecr}{\mathbf{r}}
\newcommand{\vecb}{\mathbf{b}}
\newcommand{\vaa}{\mathbf{a}_\alpha}
\newcommand{\Hgas}{{\mathcal{H}}^{{\rm gas}}}
\newcommand{\Hspin}{{\mathcal{H}}^{{\rm spin}}}
\newcommand{\sums}{\sum_{\sigma}}
\newcommand{\sumst}{\sum_{\sigma , \tau}}
\newcommand{\hm}{h_\sigma}
\newcommand{\kb}{k_{{\scriptscriptstyle B}}}
\newcommand{\sm}{s_\sigma}
\newcommand{\ros}{\rho_\sigma}
\newcommand{\ms}{m_\sigma}
\newcommand{\sn}{s_\tau}
\newcommand{\sisi}{\sigma  \sigma}
\newcommand{\st}{\sigma  \tau}
\newcommand{\dd}{{\scriptstyle{2 2}}}
\newcommand{\uu}{{\scriptstyle{1 1}}}
\newcommand{\ud}{{\scriptstyle{1 2}}}
\newcommand{\vst}{\varphi_{\st}}
\newcommand{\Jst}{J_{\st}}
\newcommand{\Jss}{J_{\sisi}}

\newcommand{\calH}{\mathcal{H}}
\newcommand{\B}{\mathcal{B}}
\newcommand{\vk}{\mathbf{k}}
\newcommand{\vh}{\mathbf{h}}
\newcommand{\vz}{\mathbf{0}}
\newcommand{\vm}{\mathbf{m}}
\newcommand{\vro}{\boldsymbol{\rho}}
\newcommand{\ml}{\mathbf{\Lambda}}
\newcommand{\mlk}{\mathbf{\Lambda}(\vk; \vla)}
\newcommand{\mlmu}{\mathbf{\Lambda}^{-1}}
\newcommand{\Jtst}{\widehat{J}_{\st}}
\newcommand{\Jtss}{\widehat{J}_{\sisi}}
\newcommand{\Gmn}{G_{\st}}
\newcommand{\Lm}{\Lambda_-}
\newcommand{\Lp}{\Lambda_+}
\newcommand{\Lpmk}{\Lambda_{\pm} (\vk; \vla)}
\newcommand{\LmLp}{\Lambda_- \Lambda_+}

\newcommand{\lb}{\bar{\lambda}}
\newcommand{\ld}{\lambda^\dagger}
\newcommand{\lh}{\check{\lambda}}
\newcommand{\ldh}{{\check{\lambda}^\dagger}}
\newcommand{\ldd}{\lambda^{\dagger^{{\scriptstyle{2}}}}}
\newcommand{\ldhd}{\lh^{\dagger^{{\scriptstyle{2}}}}}
\newcommand{\ldht}{\lh^{\dagger^{{\scriptstyle{3}}}}}
\newcommand{\ldhq}{\lh^{\dagger^{{\scriptstyle{4}}}}}
\newcommand{\lbm}{\lambda_\sigma}

\newcommand{\dJb}{\Delta \bar{J}}
\newcommand{\dJd}{\Delta J^\dagger}
\newcommand{\dJh}{\Delta \check{J}}
\newcommand{\dJdh}{\Delta \check{J}^\dagger}
\newcommand{\dJc}{\Delta J^{\, 2}}

\newcommand{\Jtud}{\widehat{J}_{\ud}}
\newcommand{\Jtudk}{\widehat{J}_{\ud}(\vk)}
\newcommand{\Jtstk}{\widehat{J}_{\sigma \tau}(\vk)}
\newcommand{\rac}{\sqrt{\left[ \ld + \dJd (\vk) \right]^2 +
    \fuq [\Jtud (\vk)]^2 }}
\newcommand{\jz}{j_0}
\newcommand{\jzd}{{\jz}^2}
\newcommand{\mb}{\bar{m}}
\newcommand{\md}{m^{\dagger}}
\newcommand{\mdd}{m^{\dagger^{{\scriptstyle{2}}}}}
\newcommand{\mh}{\check{m}}
\newcommand{\mdh}{\check{m}^{\dagger}}
\newcommand{\mdhd}{\mh^{\dagger^{{\scriptstyle{2}}}}}
\newcommand{\hb}{\bar{h}}
\newcommand{\hd}{h^{\dagger}}
\newcommand{\hdd}{h^{\dagger^{{\scriptstyle{2}}}}}
\newcommand{\hh}{\check{h}}
\newcommand{\hdh}{\check{h}^{\dagger}}
\newcommand{\thm}{\theta_m}
\newcommand{\thh}{\theta_h}
\newcommand{\Them}{\Theta_{\min}}
\newcommand{\intk}{\int_{\vk}}

\newcommand{\NN}{{\scriptscriptstyle N N}}
\newcommand{\GNN}{G_{\NN}}
\newcommand{\ZZ}{{\scriptscriptstyle Z Z}}
\newcommand{\GZZ}{G_{\ZZ}}
\newcommand{\GNZ}{G_{\NZ}}
\newcommand{\SNN}{S_{\NN}}
\newcommand{\SZZ}{S_{\ZZ}}

\newcommand{\cO}{{\mathcal{O}}}

\newcommand{\Tc}{T_c}

\newcommand{\roc}{\rho_c}

\newcommand{\N}{{\scriptscriptstyle N}}
\newcommand{\Z}{{\scriptscriptstyle Z}}
\newcommand{\X}{{\scriptscriptstyle X}}
\newcommand{\Y}{{\scriptscriptstyle Y}}
\newcommand{\XY}{{\scriptscriptstyle X Y}}
\newcommand{\NZ}{{\scriptscriptstyle N Z}}

\newcommand{\rotot}{\rho}

\newcommand{\ppk}{\phi (\vk)}
\newcommand{\calN}{\mathcal{N}}
\newcommand{\ns}{\mathcal{S}}

\newcommand{\fud}{\mbox{$\frac{1}{2}$}}
\newcommand{\fuq}{\mbox{$\frac{1}{4}$}}
\newcommand{\fuh}{\mbox{$\frac{1}{8}$}}

\newcommand{\vir}{\, ,}
\newcommand{\pt}{\, .}
\newcommand{\dJtss}{\Delta \widehat{J}_{\sisi}}
\newcommand{\kTr}{(\vk ; T, \vro)}
\newcommand{\kl}{(\vk ; \vla)}
\newcommand{\zl}{(\vz ; \vla)}

\newcommand{\Tvh}{(T,\vh)}

\newcommand{\dst}{\delta_{\sigma , \tau}}
\newcommand{\pe}{\! = \!}
\newcommand{\pneq}{\! \neq \!}

\newcommand{\calG}{\mathcal{G}}
\newcommand{\calL}{\mathcal{L}}
\newcommand{\calLs}{\mathcal{L}_\sigma}
\newcommand{\rlddjzd}{\sqrt{\ldd + {\jz}^2}}
\newcommand{\Rud}{R_{\ud}}
\newcommand{\Rb}{\bar{R}}
\newcommand{\Rd}{R^\dagger}
\newcommand{\sym}{{\textrm{sym}}}
\newcommand{\dLm}{\delta \Lambda_-}
\newcommand{\gz}{g_0}
\newcommand{\gzt}{g_0 (\theta)}
\newcommand{\gu}{g_1}
\newcommand{\gd}{g_2}
\newcommand{\lsz}{l_{\sigma,0}}
\newcommand{\lsu}{l_{\sigma,1}}
\newcommand{\lsd}{l_{\sigma,2}}
\newcommand{\uusg}{u^{1/\gamma}}
\newcommand{\uumusg}{u^{1-1/\gamma}}
\newcommand{\ct}{c_t}
\newcommand{\cm}{c_{m}}
\newcommand{\cRz}{{\mathcal{R}^0}}
\newcommand{\Ndema}{{\mathfrak{n}}}

\newcommand{\volz}{v_0}
\newcommand{\Gt}{\widehat{G}}
\newcommand{\St}{S}
\newcommand{\Jt}{\widehat{J}}
\newcommand{\stm}{\widehat{s}_\sigma}
\newcommand{\Det}{\textrm{Det}}
\newcommand{\rme}{\textrm{e}}
\newcommand{\sext}{{\textrm{ext}}}
\newcommand{\sint}{{\textrm{int}}}
\newcommand{\cF}{\mathcal{F}}

\begin{document}

\title{Criticality in multicomponent spherical models : results and cautions}

\date{\today}

\author{Jean-No\"el Aqua}
\altaffiliation{\'Ecole centrale Marseille}
\affiliation{Institut
Matériaux Microélectronique Nanosciences de Provence, Aix-Marseille
Université, UMR 6242, 13397 Marseille, France}

\affiliation{Institut de Recherche sur les Phénomènes Hors
Équilibre, Aix-Marseille Université, UMR 6594, 13384 Marseille,
France}

\author{Michael E. Fisher}

\affiliation{Institute for Physical Science and Technology,
University of Maryland, College Park, Maryland 20742, USA}

\begin{abstract}
To enable the study of criticality in multicomponent fluids, the
standard spherical model is generalized to describe an $\ns$-species
hard core lattice gas. On introducing $\ns$ spherical constraints,
the free energy may be expressed generally in terms of an
$\ns\times\ns$ matrix describing the species interactions. For
binary systems, thermodynamic properties have simple expressions,
while all the pair correlation functions are combinations of just
two eigenmodes. When only hard-core and short-range overall
attractive interactions are present, a choice of variables relates
the behavior to that of one-component systems. Criticality occurs on
a locus terminating a coexistence surface; however, except at some
special points, an unexpected ``demagnetization effect'' suppresses
the normal divergence of susceptibilities at criticality and
distorts two-phase coexistence. This effect, unphysical for fluids,
arises from a general lack of symmetry and from the vectorial and
multicomponent character of the spherical model. Its origin can be
understood via a mean-field treatment of an XY spin system below
criticality.
\end{abstract}

\pacs{64.60.F-, 61.20.Qg, 05.50.+q, 64.70.F-}

\maketitle

\section{Introduction}
\label{intro}

Criticality in liquid-vapor or fluid-fluid phase separation still
warrants study: even after the advent of renormalization group
theory, and its successful comparisons with experiment, open
questions remain. One example is criticality in charged fluids such
as electrolytes, molten salts, ionic solutions, etc. The long range
of the Coulomb interactions impedes the application of most
established methods and the interplay between electrostatic effects
and long-range critical fluctuations is not fully understood
theoretically. Indeed, the basic issue of the universality class of
ionic fluids has been under debate for many years \cite{wein&schr01}
and some questions still remain open. To gain insight into this and
related problems, exactly soluble models can be valuable. Indeed,
even if a model needs to be considered with circumspection in light
of unavoidable simplifications, it may reveal significant features
of criticality beyond those established by scaling and
renormalization group analyses.

In the history of models in statistical mechanics, the spherical or,
equivalently, the mean spherical model
\cite{berl&kac52,lewi&wann52}, has played a special role. This
``poor man's'' Ising model \cite{mef05,jn&mefrmp} has proved to be a
mine of information because of its mathematical tractability: Thus
only as regards criticality, one can readily investigate
\cite{mef05,jn&mefrmp,joyc72} the role of dimensionality, scaling
relations, finite size effects \cite{barb&mef73}, and the influence
of long-range integrable interactions (such as $1/r^{d+\sigma}$,
where $d$ is the dimension of the space and $\sigma \! > \! 0$).
Consequently the spherical model has been applied in many physical
situations, initially ferromagnets and later spin glasses
\cite{kost&thou76}, quantum transitions \cite{tu&weic94}, spin
kinetics \cite{paes&henk03}, actively mode-locked lasers
\cite{gord&fisc04}, critical Casimir forces \cite{cham&dant04}, etc.
The model became all the more interesting when it appeared
\cite{stan68} that it belongs as a limiting case, $n\rightarrow
\infty$, to the important class of spin systems in which $n$ is the
dimension of the order parameter (with $n \pe 1$, $2$, $3$, $\cdots$
for Ising, XY, Heisenberg, $\cdots$ models).

It is natural, therefore, to consider spherical models with
long-range Coulombic coupling. A pioneering investigation of a
one-component plasma (OCP) spherical model has been undertaken by
Smith \cite{smit88}; but the limitations of an OCP model are well
known and, in particular, a gas-liquid transition and corresponding
critical behavior cannot be realized. Conversely, to treat
electrolyte solutions a realistic model should first represent the
neutral solvent, typically water; then two further species, namely,
positive and negative ions, must be accounted for. Even if the
solvent is appoximated by a uniform, structureless dielectic medium,
a colloidal system, for example, requires not only the macroions and
their microscopic counterions but also the representation at some
level of an ionic salt; thereby a ternary or quaternary system is
called for. Accordingly it is desirable to develop spherical models
for \emph{multicomponent systems}. That is the aim of this paper.
The investigation of the multicomponent model proves interesting in
itself although we will focus on the conclusions that can be drawn
for simple binary fluids with short-range attractive interactions;
applications to ionic fluids are presented elsewhere
\cite{jn&mef04PRL,jn&mef04JPA,jn&mefsfp}.

\begin{figure}
  \includegraphics[width=8cm]{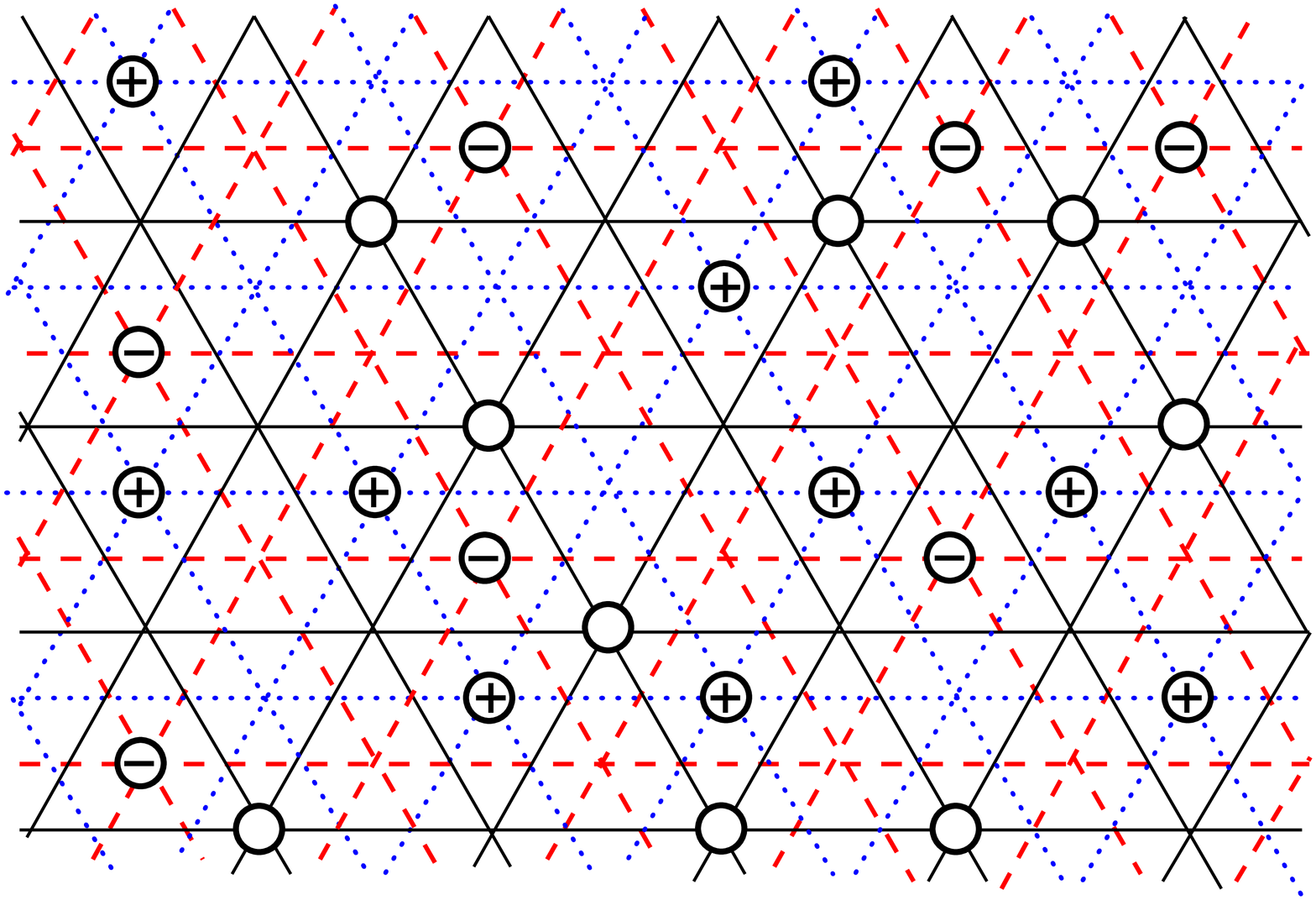}\\
  \caption{(Color online) Illustration of a two-dimensional ternary lattice gas with species
  labelled $\sigma \pe +,-,0$. Particles of each species occupy only one of
  the interlacing sublattices shown as dotted, broken, and solid lines, respectively;
  however, each particle may interact with all others via pairwize potentials
  $\phi_{\st} (\mathbf{R}^\sigma-\mathbf{R}^\tau)$.}\label{figgeom}
\end{figure}

Explicitly, we address a lattice gas with $\ns$ species of
particles, labelled $\sigma \pe 1, 2, \cdots, \ns$, in the grand
canonical ensemble. Particles of a given species may occupy or leave
vacant sites of only one sublattice so that the displacements
separating the different interlaced sublattices introduce the
crucial hard-core effects in a direct and transparent manner: see
Figure \ref{figgeom}. We will use the vectors
$\vro\pe\{\rho_\sigma\}$, $\vm$, $\vmu$, $\vh$, etc., to denote the
corresponding sets of densities, magnetizations, chemical
potentials, magnetic fields, etc., for the $\ns$ species. Using the
correspondence between lattice-gas and Ising spin models, and
enforcing the $\ns$ distinct spherical conditions with Lagrange
multipliers, we extend to this multicomponent situation, the usual
spherical model approach. This yields the free energy in terms of an
$\ns \! \times \ns$ matrix that describes the pairwise interactions
between the different species : see Section \ref{secgal} below and
Eqns.~\eqref{deffsdis}-\eqref{deffs}. It transpires that the
singular part displays a form similar not only to one-component
spherical models but also to Onsager's exact expression for the 2D
Ising model \cite{mef05,jn&mefrmp}.

In the case of binary fluids ($\ns \pe 2$), considered in Section
\ref{Deuxesp}, major simplifications allow us to obtain explicitly
all the thermodynamic and correlation properties in terms of the
eigenvalues of the interaction matrix. The density correlations or
(for charged fluids) charge correlations for the different species,
appear as \emph{combinations} of two eigenmodes. These contributions
become uncoupled only in the often considered but usually
unrealistic fully symmetric case.

To study critical behavior we go on in Section \ref{attractive} to
consider systems with only hard cores and sufficiently attractive
short-range interactions. With a proper choice of mixing
coefficients, one can define $\mh$ and $\mdh$, two linear
combinations of the species mean magnetizations or local densities,
so that the usual spherical model critical singularities occur in
the $(T,\mh)$ plane at fixed $\mdh$. Thence, a critical locus
emerges in the full $(T,\mh,\mdh)$ space which, together with a
first-order surface, describes the influence of composition on the
location of phase separation in the system. Via standard geometric
arguments, criticality in the multicomponent model can then be
deduced from corresponding one-component systems. Precisely, the
same critical universality classes are realized as in short-range
(attractive) spherical models.

Nevertheless, a significant difference arises in the \emph{equation
of state} where a new, unexpected mixing term appears. This can be
understood heuristically as a type of demagnetization effect arising
as a consequence of the vectorial character of the model coupled to
asymmetry and multispecies features. This term indeed suppresses the
normal divergence of susceptibilities at criticality; furthermore,
it induces a linear dependence of the chemical potential and
pressure as functions of the total density in the two-phase region!
These are certainly undesirable and unphysical features of any fluid
model. This  wayward behavior reinforces the remark \cite{mef92}
that, because of the \emph{de facto} vectorial character of the
order parameter in spherical models, their predictions must be
handled with perspicacity when modeling fluids.

To gain some further insight into this unanticipated
``demagnetization effect'' we study in Section \ref{model} an XY
model beneath $\Tc$ using a mean-field approach in which the
vectorial character of the order parameter, coupled to an asymmetry
of the external fields, leads transparently to a very similar
demagnetization effect. Finally, some general conclusions are drawn
in Section \ref{conc}.


\section{Multicomponent spherical models}
\label{secgal}

\subsection{Fluid and spin systems}
We consider a $d$-dimensional lattice fluid in the grand-canonical
ensemble that consists of $\ns$ species labeled $\sigma = 1, \cdots,
\ns$, each being associated explicitly with only one of $\ns$
identical interlaced sublattices (see Fig. \ref{figgeom}). Every
sublattice is taken as the image of a periodic reference sublattice
$\cRz$ after translation by a vector $\vds$ so that every site $i$
on a lattice $\sigma$ is characterized by a position $\vRis = \vRiz
+ \vds$. The reference sublattice is generated by the vectors $\vaa$
($\alpha \pe 1, \cdots, d$), has a unit cell volume $\volz$ and
contains $\calN \pe \prod_{\alpha} N_\alpha$ sites at positions
$\vRiz \pe \sum_\alpha \vRtiaz \vaa$ specified by the integers
$\vRtiaz \pe 1, 2, \cdots, N_\alpha$.

It is well-known that a grand-canonical lattice fluid is in
correspondence with a canonical spin system \cite{lee&yang52s}.
Indeed, let us write the grand partition function of a fluid as
\begin{equation}
  \label{Xi}
  \Xi (T, \vmu) = \prod_\sigma \left(\sum_{N_\sigma} \frac{1}{N_\sigma !}
  \sum_{\vecr_i^\sigma} \right)
  \exp \left[- \beta (\Hgas - \sum\nolimits_\tau \mu_\tau N_\tau) \right] ,
\end{equation}
where $\beta$ is the inverse temperature $1/\kb T$, while $N_\sigma$
and $\mu_\sigma$ denote the number of particles and chemical
potential of species $\sigma$. The Hamiltonian $\Hgas$ is expressed
as a sum over particles $k\pe 1, \cdots, n^\sigma$ and $l\pe 1,
\cdots, n^\tau$ as
\begin{equation}
  \label{}
  \Hgas = \fud \sum_{(\sigma,k) \neq (\tau,l)} \vst(\vecr^\sigma_k - \vecr^\tau_l) \vir
\end{equation}
where $\vst$ is the pair interaction potential while
$\vecr^\sigma_k$ is the position of the $k$-th particle of species
$\sigma$, occupying sites on the lattice $\sigma$. Considering a
system with hard cores, i.e., $\varphi_{\sisi} (\vz) \pe + \infty$,
the sum in \eqref{Xi} refers to configurations where the local
lattice density $n_\sigma (\vRis) = \sum_{k} \delta(\vRis -
\vecr^\sigma_k)$ can be only $0$ or $1$, so that the local spin
variable
\begin{equation}
  \label{}
  \sm (\vRis) = 2 n_\sigma (\vRis) - 1 \vir
\end{equation}
takes the values $\pm 1$ as in the Ising model. A straightforward
generalization of the procedure described in \cite{lee&yang52s} then
leads to the partition function of a spin system, namely,
\begin{equation}
  \label{xispin}
  \Xi (T, \vmu)
  = \prod_\sigma \sum_{\sm(\vRis) = \pm 1} \exp \left(- \beta \, \Hspin \right) ,
\end{equation}
where the spin Hamiltonian is
\begin{equation}
  \label{hspin}
  \Hspin = - \fud \sum_{\substack{(\sigma,i),(\tau,j)}}
   \Jst (\vRis - \vRjt) \sm (\vRis) \sn(\vRjt) 
    - \sums h_\sigma \sum_{i} \sm (\vRis) + \mathcal{H}_o (\vmu) \pt
\end{equation}
In this correspondence, the link between the coupling energies and
pair interactions is
\begin{equation}
  \label{}
   \Jst (\vRis - \vRjt) = - \fuq \, \vst (\vRis - \vRjt) \quad
   \textrm{if \, $\vRis \neq \vRjt$ ,}
\end{equation}
with $\Jss (\vz) \pe \vz$, while the external fields are given by
\begin{equation}
  \label{hs2mus}
  h_\sigma = \fud \mu_\sigma - \fuq \phi_\sigma  ,
\end{equation}
with reference level
\begin{equation}
\label{defphiz}
 \phi_\sigma = \sum_{(\tau,j) \neq
(\sigma,i_0)}
   \varphi_{\sigma \tau} (\mathbf{R}_{i_0}^\sigma - \vRjt) \vir
\end{equation}
where $i_0$ is a fixed position. Finally, the background term in
\eqref{hspin} is merely
\begin{equation}
  \label{defHz}
   \mathcal{H}_o (\vmu)  = - \fud \calN \sum_\sigma \mu_\sigma + \fuh
    \sum_{(\sigma,i) \neq (\tau,j)} \vst (\vRis - \vRjt) \pt
\end{equation}

The correspondence between fluid and spin systems follows
straightforwardly for the other properties. For instance, the local
density $\ros (\vecr) \pe n_\sigma(\vecr) / \volz$ is related to the
local spin via
\begin{equation}
  \label{}
  \volz \ros (\vRis) = \fud \left[ s_\sigma
  (\vRis) + 1 \right] \vir
\end{equation}
while the species correlation functions
\begin{equation}
  \label{}
  \Gmn (\vRis - \vRjt; T, \vmu)  =  
  \langle
  \rho_\sigma (\vRis) \rho_\tau (\vRjt) \rangle
  - \langle \rho_\sigma (\vRis) \rangle  \langle
\rho_\tau (\vRjt) \rangle \vir
\end{equation}
are related to spin correlations via
\begin{equation}
  \label{Gmn}
  \volz^2 \Gmn (\vRis - \vRjt; T, \vmu) = 
  \fuq \left[\langle \sm (\vRis) \sn (\vRjt)   \rangle - \langle \sm (\vRis)
    \rangle \langle \sn(\vRjt) \rangle \right]   \pt
\end{equation}
As usual, the angular brackets denote grand-canonical expectation
values.

In order to define density or charge correlations simply in this
lattice geometry, it is convenient to work in Fourier space. We
consider periodic boundary conditions and define Fourier series with
respect to the reference sublattice $\cRz$ by
\begin{equation}
  \label{}
  \stm (\vk) = \sum_i \rme^{-i \vk \cdot \vRis} \sm(\vRis) .
\end{equation}
Then, when $\Jst$ is periodic over the reference sublattice, we may
write
\begin{equation}
  \label{deftfJst}
  \Jt_{\st} (\vk) = \sum_i \rme^{-i \vk \cdot (\vRis -
  \mathbf{R}^{\tau}_{j_0})}
  J_{\sigma\tau} (\vRis - \mathbf{R}^{\tau}_{j_0}) 
 = \sum_j \rme^{-i \vk \cdot (\mathbf{R}^{\sigma}_{i_0} - \vRjt)}
 J_{\sigma\tau} (\mathbf{R}^{\sigma}_{i_0} - \vRjt)  \vir
\end{equation}
with any fixed positions $i_0$ and $j_0$. The wave vectors should be
combinations of the reciprocal vectors $\vecb_\alpha$ (defined by
$\vaa \mathbf{\cdot} \vecb_{\alpha'} \pe 2 \pi
\delta_{\alpha,\alpha'}$) such as $\vk \pe \sum_\alpha
\tilde{k}_\alpha \vecb_\alpha$ with $\tilde{k}_\alpha \pe 0, \pm
1/N_\alpha, \pm 2/N_\alpha, \cdots$. In the following, the first
Brillouin zone is denoted as $\B$. The density correlation function
$\GNN$, and, for fluids of particles carrying charges $q_\sigma$,
charge $\GZZ$ and charge-density $\GNZ$ correlations can then be
defined via
\begin{equation}
  \label{defGNN}
  \Gt_{\XY} \kTr = \sumst q_\sigma^{\vartheta_{\X}} q_\tau^{\vartheta_{\Y}}
  \Gt_{\sigma \tau} \kTr \vir
\end{equation}
where $X$ and $Y$ stand either for $N$ or $Z$, with $\vartheta_{\N}
= 0$, $\vartheta_{\Z} = 1$. We also define structure factors as
\begin{equation}
\label{defSXY}
  \St_{\XY} \kTr = \frac{\volz}{\rotot \, q^{\vartheta_{\X}+\vartheta_{\Y}}}
    \Gt_{\XY} \kTr  \vir
\end{equation}
where $q$ is an elementary charge, while the total density is
$\rotot \pe \sum_\sigma \! \langle \ros \rangle$. The term $\volz$
compensates here for the homogeneity difference between the discrete
and continuum Fourier transforms.


\subsection{Mean spherical model}

We are not able to perform the multiple sums in \eqref{xispin} in
general. Instead, we adopt the appropriate mean-spherical model
\cite{lewi&wann52} and compute the multiple integral
\begin{equation}
  \label{Xip}
  \Xi' (T, \vh) = \int \prod_\sigma  d\sm \, \rme^{- \beta \, \calH'}
  \vir
\end{equation}
with
\begin{equation}
  \label{defHp}
  \calH' = \Hspin + \sums \lds \sum_i \sm^2 (\vRis) \pt
\end{equation}
As usual, the Lagrange multipliers $\lds$ are introduced to allow
imposition of the mean spherical conditions which need to be
enforced uniformally for every species; specifically, the relations
\begin{equation}
  \label{csi}
  \bigl< \sum_i  \sm^2(\vRis) \bigr> = \calN \quad , \quad
  \sigma = 1 , \cdots \ns \pt
\end{equation}
define the Lagrange multipliers or \emph{spherical fields} $\lds$ as
implicit functions of $(T, \vh)$. Consequently, the free energy per
site (of the reference sublattice) is
\begin{equation}
  \label{}
  - \beta f \bigl[ T, \vh, \vla(T,\vh)
  \bigr] = \ln \Xi' (T, \vh) / \calN   \vir
\end{equation}
in terms of which the spherical conditions \eqref{csi} can be
rewritten as
\begin{equation}
  \label{csgal}
  \langle \sm^2 \rangle = \left. \frac{\partial f}{\partial \lds} \right|_{T,
    \vh, {\tilde{\lambda}_{\tau}}; \tau \neq \sigma} = 1 \quad , \quad
  \sigma = 1, \ldots \ns \pt
\end{equation}
As already remarked in Sec.~\ref{intro}, the spherical model in this
form describes exactly spin models with fixed length or continuous
$n$-component spins in the limit $n \rightarrow \infty$ with
appropriate scalings \cite{stan68,sarb&mef78}.

As standard \cite{lewi&wann52}, the calculation of $\Xi'$ is
performed in Fourier space. For consistency, we will suppose that
$\Jst$ satisfies the symmetry condition $\Jst (\vRis - \vRjt) \pe
\Jst \left[ - (\vRis - \vRjt) \right]$. The calculation is then a
straightforward generalization of the mean-spherical techniques used
for single-species systems. The free energy per site can be
decomposed into a sum of three parts: $f \pe f_s + f_h + f_o$. The
singular part of the free energy is
\begin{equation}
  \label{deffsdis}
   - \beta f_s ( T, \vh) = - \frac{1}{2 \calN} \sum_{\vk \in \B}
   \ln \left\{ \beta^{\ns} \Det \left[ \mlk \right] \right\} .
\end{equation}
where the sum runs over the reference Brillouin zone $\B$ while
$\mlk$ is the $\ns \times \ns$ interaction matrix with elements
\begin{equation}
\label{defml} \Lambda_{\st} (\vk; \vla) = \dst \bigl[ \lambda_\sigma
+ \dJtss (\vk) \bigr] - \fud (1- \dst) \Jtstk \vir
\end{equation}
in which for any function $\widehat{g} (\vk)$ we employ the notation
\begin{equation}
  \label{dJmn}
 \Delta \widehat{g} (\vk) = \fud \left[ \widehat{g}(\vz) - \widehat{g}(\vk) \right] \vir
\end{equation}
while, dropping the tildes in \eqref{defHp} and \eqref{csgal}, the
shifted or net spherical fields are
\begin{equation}
  \label{}
  \lbm = \lds - \fud \Jtss (\vz) \pt
\end{equation}
On the other hand, the $h$-dependent part of the free energy is
given by
\begin{equation}
  \label{deffh}
   - \beta f_h ( T, \vh) = \fuq \beta  \langle \vh | \mlmu (\vz,\vla) | \vh
   \rangle \vir
\end{equation}
while the analytic background part, following from \eqref{defHz}, is
\begin{equation}
  \label{deffz}
   - \beta f_o ( T, \vh) = \fud \ns \ln \pi  - \beta
   \mathcal{H}_o / \calN \vir
\end{equation}
which will be neglected henceforth. Because of the logarithm in
\eqref{deffsdis}, these results are valid while the eigenvalues of
the matrices $\mlk$ are positive for every $\vk$; when one vanishes,
the expressions \eqref{deffsdis} and \eqref{deffh} become singular
and phase transitions are implicated.

The last step is taking the thermodynamic limit $\calN \!
\rightarrow \! \infty$ (valid provided the Fourier transforms remain
well-defined) with the result
\begin{equation}
  \label{deffs}
   - \beta f_s ( T, \vh) = - \fud \intk
   \ln \left\{ \beta^{\ns} \Det \left[ \mlk \right] \right\}  \vir
\end{equation}
where $\intk$ is a short-hand notation for $\int_{\vk \in \B} \volz
d^d \vk / (2 \pi)^d $, while the $\vh$-dependent free energy $f_h$
is still given by \eqref{deffh}. At this point, it is worth noting
that the structure of $f_s$, as an integral over the Brillouin zone
of the logarithm of the interactions in Fourier space, is similar to
that present in Onsager's exact solution of the 2D Ising model
\cite{onsa44}. The consequences for charged systems are dramatic
\cite{jn&mef04PRL,jn&mef04JPA,jn&mefsfp} since this form determines
the coupling or decoupling of correlations in symmetric and
asymmetric systems.

With these results in hand, we find that the mean particle densities
$\ros = \langle \ros (\vRis) \rangle$ are related to the mean
magnetizations via
\begin{equation}
  \label{defms}
  2 \ros \volz - 1 = \ms = \langle \sm \rangle = - \left. \frac{\partial f_h}{\partial \hm}
  \right|_{T, h_\tau, \vla; \tau \neq \sigma} \vir
\end{equation}
which, in turn, enter the free energy in standard manner as
\begin{equation}
  \label{}
  f_h = - \fud \sums \ms \hm \pt
\end{equation}
As a result of \eqref{deffh} and \eqref{defms}, the link between
$\vh$ and $\vm$ is then merely
\begin{equation}
  \label{hfnm}
  \vh = \fud \vmu - \fuq \boldsymbol{\phi} = 2 \ml (\vz) \vm ,
\end{equation}
where $\vmu \pe \{ \mu_\sigma\}$ and we have recalled \eqref{hs2mus}
and introduced a fixed vector $\boldsymbol{\phi} \pe \{\phi_\sigma
\}$: see \eqref{defphiz}. Finally, in the thermodynamic limit, the
spin-spin correlation functions are given by
\begin{equation}
  \label{Gss}
  \langle \sm (\vRis) \sn (\vRjt) \rangle = - \frac{1}{2-\delta_{\sigma,\tau}}
  \left. \frac{\partial (f-f_o)}{\partial
\Jst (\vRis - \vRjt)} \right|_{T, \vh, \vla}  \vir
\end{equation}
where $2-\delta_{\sigma,\tau}$ is merely a symmetry factor, while we
recall \eqref{Gmn} for the density correlation functions $G_{\st}$.


\section{Binary systems}
\label{Deuxesp}

The previous analysis holds for an arbitrary number of species. From
here on, however, we focus on the simplest case, i.e., binary
mixtures with species labels $1$ and $2$. For many properties, it is
useful  to decompose densities, chemical potentials, etc., in terms
of means and differences; so for every function $g_\sigma$ (or
$g_{\sisi}$) we define
\begin{equation}
  \label{defbd}
  \bar{g} = \fud \left( g_1 + g_2 \right) , \quad \quad g^\dagger = \fud \left( g_1 -
    g_2 \right) \pt
\end{equation}
Moreover, for simplicity, we suppose that the translation vectors
$\vds \pe \sum_\alpha \tilde{\delta}_\alpha \vaa$ satisfy
$\tilde{\delta}_\alpha \pe 0$ or $1/2$ so that the Fourier
transforms $\Jt_{\st}$ are real.

\subsection{Basic features}

For the case $\ns \pe 2$, simplifications allow more explicit
results. First, let us introduce the energy scale
\begin{equation}
  \label{defjz}
  \jz = \fud \Jtud (\vz) \vir
\end{equation}
and, following \eqref{defbd}, write
\begin{subequations}
\begin{gather}
\label{defdJb} \dJb (\vk) = \fud \left( \Delta J_{\uu} + \Delta
J_{\dd} \right) 
, \\
\label{defdJd} \dJd (\vk) = \fud \left( \Delta J_{\uu} - \Delta
J_{\dd} \right) 
.
\end{gather}
\end{subequations}
Then the eigenvalues of the $2\times 2$ matrix
$\boldsymbol{\Lambda}$ may be written
\begin{equation}
  \label{lpm}
  \Lpmk = \lb + \dJb (\vk) \pm D \kl \vir
\end{equation}
where
\begin{equation}
D \kl = \rac \geq 0 .
\end{equation}
As remarked above, these expressions are valid when $\Lm$ and $\Lp$
are nonnegative while singularities arise only when $\Lm \kl \,
[\leq \Lp \kl] \rightarrow 0$.

Now the argument of the free energy integral in \eqref{deffs} is
$\ln \{ \beta^2 \Det [\ml]\}$, where the determinant of the
interaction matrix can now be written
\begin{equation}
\label{lmlp}
    \LmLp \kl  = u + 2\lb \dJb (\vk) - 2 \ld \dJd (\vk) + \dJc (\vk) ,
\end{equation}
where we have introduced the crucial parameter
\begin{equation}
    \label{defu}
        u (\vla) \equiv \LmLp (\vz;\vla) = \lb^2 - \ldd - \jzd  \vir
\end{equation}
which vanishes when $\Det [\ml(\vk)]$ vanishes at $\vk \pe \vz$,
while the squared interaction term in \eqref{lmlp}, namely,
\begin{equation}\label{}
    \dJc (\vk) = \jzd - \fuq {\Jtud}^{\, 2} (\vk) + \Delta \Jt_{\uu} (\vk) \Delta
    \Jt_{\dd}  (\vk) ,
\end{equation}
vanishes as $|\vk^2|$.

In terms of the eigenvalues, the spherical conditions \eqref{csgal}
become
\begin{equation}\label{cs}
    1 =  \fud \kb T \intk \frac{\lambda_\sigma + \dJtss (\vk)}{\LmLp \kl}
    + m_{\tau}^2 \vir \quad \tau \neq \sigma .
\end{equation}
Finally, the $h$-dependent part of the free energy entails
\begin{equation}
  \label{}
  \mlmu (\vz;\vla) = \frac{1}{\LmLp (\vz; \vla)}
\begin{pmatrix}
\lambda_2 \, \,& \jz \\
 \jz   \, \, & \lambda_1
\end{pmatrix} \vir
\end{equation}
while the magnetization-field or density-chemical potential relation
\eqref{hfnm} becomes
\begin{equation}\label{hs2ms}
   \fud h_\sigma =  m_\sigma  \lambda_\sigma - m_{\tau}  \jz  \vir
   \quad \tau \neq \sigma .
\end{equation}
At this point, Eqs.~\eqref{defjz}-\eqref{hs2ms} entirely define the
system and the need is to analyze their structure and consequences.


\subsection{Correlation functions}

The density pair correlation functions are given generally by
\eqref{Gmn} and \eqref{Gss} which, when $\ns \pe 2$, reduce to
\begin{gather}
\label{gmm}
  \Gt_{\sisi} (\vk;\vla) = \frac{\kb T}{8 \volz^2} \frac{\lambda_\sigma + \dJtss (\vk)}{\LmLp \kl}
  , \quad (\sigma =1,2) ,\\
  \label{gud}
  \Gt_{\ud} (\vk;\vla) = \frac{\kb T}{16 \volz^2} \frac{\Jtudk}{\LmLp \kl}
  = \Gt_{{\scriptstyle{2 1}}} (\vk;\vla)  .
\end{gather}
In terms of these, one can  use \eqref{defGNN} to obtain the overall
density-density correlation function, $\Gt_{\NN}$ and, the
complementary compositional correlations or, for charged systems,
the charge-charge correlation function $\Gt_{\ZZ}$.

From a purely mathematical perspective, it is also instructive to
decompose the fluctuations with respect to the eigenvectors of $\ml$
which, of course, depend on the wavevector $\vk$ and the fields
$\vla$. Thus if we define $\phi(\vk)$ via
\begin{equation}
  \label{}
  \tan \ppk = 2 \left\{ D \kl
   - \bigl[ \ld +\dJd (\vk) \bigr] \right\} / \Jtudk ,
\end{equation}
it can be interpreted as the angle determined by the eigenvector
associated with $\Lp \kl$ relative to the $\sigma \pe 1$ axis. Then
if we introduce the density fluctuations $\rho^+ \kl$ and $\rho^-
\kl$ via
\begin{equation}
\label{combin}
    \rho^\pm =
    \left[ \rho_1 \cos \phi \kl \mp \rho_2 \sin
    \phi \kl   \right] /\sqrt{2} ,
\end{equation}
and define the corresponding correlation function $\Gt_{\pm \pm}$ in
the natural way, we find
\begin{subequations}
  \begin{gather}
    \label{grp}
      \Gt_{+ +} (\vk; \vla)
       = \frac{\kb T}{16 \volz^2} \frac{1}{\Lp \kl} \vir \\
    \label{grpd}
         \Gt_{- -} (\vk; \vla)
      = \frac{\kb T}{16  \volz^2} \frac{1}{\Lm \kl} \pt
  \end{gather}
\end{subequations}
while $\Gt_{+ -} \pe \Gt_{- +}$ vanishes identically.

However, the eigenmodes \eqref{combin} will rarely be of direct
physical significance. Rather the physically accessible
fluctuations, represented in particular by the structure functions
introduced in \eqref{defGNN} and \eqref{defSXY}, will typically
involve a \emph{mixture} of the underlying eigenmodes. Specifically
we find
\begin{subequations}
\label{natdec}
  \begin{equation}
    \label{gnndecomp}
    \frac{\St_{\NN} \kTr}{\kb T / 4 \rho \volz}  =
     \frac{B \kl}{\Lm \kl} +   \frac{1 - B \kl }{\Lp \kl}  \vir \\
  \end{equation}
and, for charged systems with $q_+ \pe - q_- \pe q$,
  \begin{equation}
    \label{gzzdecomp}
    \frac{\St_{\ZZ} \kTr}{\kb T / 4 \rho \volz} =   \frac{B \kl}{\Lp \kl} +
      \frac{1-B \kl}{\Lm \kl} \vir
  \end{equation}
\end{subequations}
where the mixing amplitude $B$ is
\begin{equation}
  \label{defB}
  B \kl =  \fud + \fuq \Jtudk / D \kl \pt
\end{equation}
Evidently, singular behavior, anticipated at criticality in $\Lm
\kl$, will in general affect \emph{both} $\SNN$ and $\SZZ$ as we
discuss in detail elsewhere
\cite{jn&mef04PRL,jn&mef04JPA,jn&mefsfp}.

However, a special situation arises when the two species 1 and 2 are
symmetrically related so that $\Jt_{\uu} (\vk) \pe \Jt_{\dd} (\vk)$
which implies, via \eqref{defdJd}, $\dJd(\vk) \equiv 0$. For a
charged system this corresponds to complete charge symmetry as
exemplified most simply in the \emph{restricted primitive model}
(RPM) of equisized hard spheres with charges of equal magnitude but
opposite sign. But neutral systems where species 1 and 2 differ only
in chirality demand a symmetric description quite naturally. Then,
on the locus of symmetry where $\rho_1 \pe \rho_2$ (corresponding to
electroneutrality in 1:1 ionic fluids) one has $\mu_1 \pe \mu_2$
and, hence, via \eqref{hfnm}, $\lambda_1 \pe \lambda_2$ and thence,
via \eqref{defbd} $\ld \equiv 0$. In this case one sees from
\eqref{defB} that $B \kl$ vanishes identically so that the
eigenmodes precisely specify $\SNN$ and $\SZZ$ which, therefore,
become totally \emph{decoupled}! This turns out to play a crucial
role in the study of charge screening near ionic criticality
\cite{jn&mef04PRL,jn&mef04JPA,jn&mefsfp} albeit for generally
unrealistic charge-symmetric systems.

A small technical detail deserves mentioning in this fully symmetric
case if $\Jt_{\ud} (\vk)$ should change sign for $\vk \pneq \vz$
(which is not unreasonable); then the ratio $\Jt_{\ud}/D\kl$ in
\eqref{defB} together with $\Lp$ and $\Lm$ involve nonanalytic
absolute values but in such a way that the combinations $\St_{\NN}$
and $\St_{\ZZ}$ in \eqref{natdec} remain completely analytic.


Finally, the cross charge-density structure function is also
expressible as a combination of the two eigenmodes via
\begin{equation}
  \label{gnzlmlp}
  \frac{\St_{\NZ} \kTr}{\kb T / 8 \volz \rotot} =  \frac{\ld + \dJd (\vk)}{D \kl}
  \left[ \frac{1}{\Lm \kl} - \frac{1}{\Lp \kl} \right] \pt
\end{equation}
As is to be anticipated, this vanishes identically on the symmetry
locus when (1,\,2) symmetry is present.

\subsection{Appropriately mixed thermodynamic variables}

Depending on the symmetry of the system, the previous relations may
be handled more or less conveniently. In the general asymmetric case
($\Jt_{\uu} \neq \Jt_{\dd}$), the spherical constraints \eqref{cs}
can be rewritten as
\begin{equation}
  \label{csl}
  1 =  \fud \kb T \intk \frac{\lb + \dJb (\vk)}{\LmLp \kl} + m^2 + {\md}^2 \vir
\end{equation}
and
\begin{equation}
  \label{csld}
  2 m \md =  \fud \kb T \intk \frac{\ld + \dJd (\vk)}{\LmLp \kl}
  \vir
\end{equation}
while the external fields are given by
\begin{subequations}
  \begin{align}
    \label{hmmd}
    \hb = & 2 \left[ m (\lb - \jz) + \md \ld \right] \vir\\
    \label{hdmmd}
    \hd = & 2 \left[ m \ld + \md (\lb + \jz)  \right]  \pt
  \end{align}
\end{subequations}
Note that, since $\Lm$ and $\Lp$ are nonnegative, the condition
\eqref{csl} is consistent with the expectations $|m| \! \leq \! 1$
and $|\md| \!\leq \!1$.

For further analysis it is convenient to introduce the basic
integral functions
\begin{gather}
  \label{defg}
    \calG (\vla) = \fud  \intk \frac{1}{\LmLp \kl} \vir \\
 \label{defls}
    \calLs (\vla) = \fud  \intk \frac{\dJtss(\vk)}{\LmLp \kl} \vir
\end{gather}
which are simple generalizations of the typical integrals involved
in the analysis of standard spherical models.

Now, in the fully symmetric case, $\dJd$, $\ld$, and $\md$ vanish
identically so that the relations \eqref{csld} and \eqref{hdmmd}
have no role to play. Then \eqref{csl} and \eqref{hmmd} closely
resemble the basic expressions for the single-species ($\ns \pe 1$)
or standard spherical model. These in turn lead to the basic
\textit{equation of state} which, in terms of the reduced
temperature variable
\begin{equation}
\label{deft} t=(T-\Tc)/\Tc ,
\end{equation}
can be written most transparently near the critical point ($T\pe
\Tc$, $m \pe 0$) as \cite{mef05,jn&mefrmp,joyc72}
\begin{equation}
\label{eqgenms}
 p_0 u^{1/\gamma} \approx c_t t + c_m m^2 ,
\end{equation}
where, recalling \eqref{defu}, $u \pe \lb^2-\jz^2$ while $\gamma \!
\geq \! 1$ is the fundamental dimensionality-dependent exponent, and
$p_0$, $c_t$, and $c_m$ are fixed positive coefficients.

Now physicochemical insight into the behavior of binary fluid
mixtures suggests strongly that their critical behavior will, when
expressed in terms of suitable density and field variables be
essentially the same as for a single-component fluid. However, the
``suitable'' or ``appropriate'' variables will, in leading order, be
linear combinations or mixtures of the related binary thermodynamic
variables, specifically, the fields and densities. Furthermore, the
appropriate mixing coefficients must, in general, be
\textit{nontrivial} functions of the state variables.

It follows that our primary task now is to find what the appropriate
mixing coefficients are. To that end, we introduce the general
linear combinations
\begin{equation} \label{deflhg}
    \lh  = \fud \left( \theta^{-1} \lambda_1 + \theta \lambda_2
    \right) , \quad \ldh = \fud \left( \theta^{-1} \lambda_1 -
    \theta \lambda_2 \right) ,
\end{equation}
together with corresponding remixed interactions
\begin{equation}
\label{defdjt}
    \dJh , \, \dJdh = \fud \left( \theta^{-1} \Delta J_{\uu}
    \pm \theta \Delta J_{\dd}  \right) .
\end{equation}
The mixing parameter $\theta$ is to be determined later. In terms of
these new variables and interactions, the basic determinant becomes
\begin{equation}
    \label{lmlpfnt}
    \Det [\ml] = u + 2 \lh \dJh (\vk) - 2 \ldh \dJdh (\vk) + \dJc (\vk) ,
\end{equation}
where, following \eqref{defu}, the value at $\vk \pe \vz$ is now
\begin{equation}
    \label{uatc}
    u (\lambda_1,\lambda_2)= \lh^2 - \ldhd - \jzd .
\end{equation}
Then it proves necessary to introduce a second state-dependent
mixing parameter $\thm$ by writing
\begin{equation} \label{defmt}
    \mh , \mdh = \fud \left(\thm^{-1} m_1 \pm \thm m_2
    \right) .
\end{equation}
In terms of these new variables and the integrals \eqref{defg} and
\eqref{defls}, the original spherical conditions \eqref{cs} become
\begin{subequations}
\label{cslt}
\begin{gather}
    \label{cslu}
  1 = \theta^{-1} \kb T (\lh - \ldh) \calG (\vla)  + \calL_2(\vla) \kb T 
  \nonumber \\+ \thm^2 (\mh^2 + \mdhd + 2 \mh \mdh ) , \\
    \label{cslde}
  1 = \theta \kb T (\lh + \ldh) \calG (\vla)  + \calL_1(\vla) \kb T  
 \nonumber \\+ \thm^{-2} (\mh^2 + \mdhd - 2 \mh \mdh ) .
\end{gather}
\end{subequations}
Finally, it is helpful to define new external fields via
\begin{equation}
\label{h2hs} \hh  = \fud \left( \thh^{-1} h_1 + \thh h_2 \right) ,
\quad \hdh = \fud \left( \thh^{-1} h_1 - \thh h_2 \right) ,
\end{equation}
which are linked to the generalized magnetizations via
\begin{multline}
    \label{ht2mt}
        \hh = \mh \left[ (\lh - \jz) \, \omega_{+}   +
        \ldh  \omega_{-} + \jz \pi^{+}_m  \right] 
         \\
         + \mdh \left[ (\lh - \jz) \, \omega_{-} +
        \ldh \omega_{+} + \jz  \pi^{+}_{m^\dagger} \right] \vir
    \mbox{\hspace{2cm}}
\end{multline}
and similarly for $\hdh$ in terms of $\pi_m^-$ and
$\pi_{m^\dagger}^-$ with $\omega_+$ and $\omega_-$ interchanged
while the coefficients $\omega_\pm$, $\pi_m^\pm$, and
$\pi_{m^\dagger}^\pm$ are found to be
\begin{gather}
    \label{defopm}
    \omega_{\pm} = \theta \thm / \thh
    \pm \, \thh/ \theta \thm \vir \\
    \pi^{\pm}_m / (\thm-1/\theta \thm) =
    \pi^{\pm}_{m^\dagger} / (\thm+1/\theta \thm)  
    \nonumber \\ = \theta/\thh \mp \, \thh \pt
\end{gather}

With these new fields and magnetizations, the free energy per site
reduces to
\begin{multline}
    \label{deffhgal}
    f_h = - \fud \left(\thh \thm + 1/\thh \thm \right) (\mh \hh + \mdh \hdh) 
    \\ + \fud \left(\thh \thm - 1/\thh \thm \right) (\mh \hdh + \mdh \hh)
    . \hspace{3cm}
\end{multline}
As we will show, the choice of the coefficients $\theta$, $\thm$ and
$\thh$ will be dictated by physical arguments in order to ensure
compact and familiar expressions for the critical behavior.


\section{Binary lattice gases with short-range attractive interactions}
\label{attractive}

To obtain explicit results for critical behavior we focus now on
binary systems with short-range interactions (in addition to the
hard cores already accounted for). Accordingly, we suppose that the
small-$\vk$ expansions of the interactions in $\vk$ space are
\begin{equation}\label{dljst}
    \Jtst (\vk) = \Jtst (\vz) \left[ 1 - k^2 \, R_{\st}^2 + \cO
    (k^4) \right] , \quad (\sigma, \tau  =1, 2) ,
\end{equation}
with fixed range parameters $R_{\st}$. Moreover, to ensure simple
criticality, we suppose that the interactions are ``overall
attractive'', which we take to mean that $\jz$, $\Delta \Jt_{\ud}$
and $\dJb$ are real and satisfy
\begin{equation}\label{}
    \label{hjud}
    \jz = \fud \Jt_{\ud}(\vz) \! > \! 0  \, \, \,
    \textrm{and } \, \Delta |\Jt_{\ud} (\vk)|  > 0 ,
    \, \, \,  \dJb (\vk) > 0
    \quad \forall \vk \neq \vz .
\end{equation}
These conditions are easily fulfilled, as, for example, when
$J_{\ud} (\vecr) \pe J_{\ud}(-\vecr)$ while $J_{\uu} (\vecr)$ and
$J_{\dd}(\vecr)$ are positive for all $\vecr$.

\subsection{Critical loci}

To identify the singularities of the binary systems, we recall that
they are signaled by the vanishing of one of the eigenvalues of
$\ml$, which occurs first  when $\Lm \kl$ vanishes. As shown in
Appendix \ref{appA}, these singularities arise only when $(i)$ $\vk
\pe \vz$, and, thence, $(ii)$ when the spherical fields $\vla$
satisfy
\begin{equation}\label{ii}
    u (\lambda_1,\lambda_2) = 0 \vir
\end{equation}
provided the \emph{asymmetries} of the interactions are \emph{not
too extreme} in the sense that, as we suppose henceforth,
$\dJd(\vk)$ [defined in \eqref{defdJd}] satisfies the conditions
\eqref{cRd} and \eqref{hdjd}.

Now, any state of the system is specified \emph{ab initio} by the
three thermodynamic fields $(T,\mu_1,\mu_2)$ which, via
\eqref{csgal}, give  $\vla \Tvh$, and then the densities $\vm$.
However, for the location of critical points, it is more convenient
to utilize the set of variables $(T,\mh,\mdh)$ introduced in
\eqref{defmt} and then to solve for $(\lh,\ldh)$ as defined in
\eqref{deflhg}.

Next, let us choose $\thm \! > \!0$ in \eqref{defmt} so that $\mh_c$
vanishes at criticality or, in other words, take $\thm \pe
(-m_{1,c}/m_{2,c})^{1/2}$. This condition will be analyzed below in
seeking a critical point, at a given value of $\mdh$. Likewise, we
choose $\theta \! > \! 0$ in \eqref{deflhg} so that $\ldh_c =0$.
This condition then enforces the link between $\theta$ and $\mdh$,
since it implies $\theta^2 \pe \lambda_{1,c}/\lambda_{2,c}$. As
established in Appendix \ref{appA}, the singularities are
characterized by $u \pe 0$, which via \eqref{uatc} means $\lh_c =
\jz$ and thence, $\lambda_{1,c} = \theta \jz$ and $\lambda_{2,c} =
\jz / \theta$.

At this point, one must pay attention to the behavior of integral
expressions \eqref{defg} and \eqref{defls} when $\Lambda_-(\vz)$
approaches zero. If we accept \eqref{dljst} we find that $\Lambda_-$
varies as $k^2$ when $\Lambda_-(\vz) \pe 0$ and then $\calG(\vla)$
and $\calLs(\vla)$ remain finite at the singularity provided $d\! >
\! d_< \pe 2$ (in this case) as seen in \cite{barb&mef91}.
Accordingly, from here on we suppose the dimensionality exceeds
$d\pe 2$ and may then write
\begin{equation}\label{}
    \calG (\vla_c) = \gzt /\jz^2 , \quad
    \calLs (\vla_c) = \gzt \lsz (\theta) /\jz  , \quad (\sigma = 1,2) .
\end{equation}
The residual $\theta$-dependence arises from $\dJh$ and $\dJdh$, see
\eqref{defdjt}, \eqref{lmlpfnt}, \eqref{defg} and \eqref{defls}.

Putting these considerations together we find that the critical
locus, $\Tc(\mdh)$ with $\mh_c \pe 0$, may be defined parametrically
via
\begin{gather}
    \label{Tc2md}
    \kb \Tc (\mdh) =  \frac{\jz ( 1-\thm^2 \mdhd)}{\gzt[1 / \theta +
    l_{2,0}(\theta)]} \vir \\
    \label{md2th}
    \mdhd =  \frac{(\theta - 1/ \theta)  +
    l_{1,0}(\theta) - l_{2,0}(\theta)}{(\theta \thm^2-1/\theta \thm^2)
    + l_{1,0}(\theta) \thm^2 - l_{2,0}(\theta)/\thm^2} \pt
\end{gather}
In fact, the latter relation must be seen as an implicit equation
giving $\theta$ as a function of $\mdh$ while, as shown below,
$\thm$ will also be related to $\theta$. Hence, if one realizes that
$\mdh$ (i.e., some combination of the densities other than the total
density) characterizes the composition of the system, the function
$T_c (\mdh)$ describes naturally the composition dependence of
criticality in the binary fluids.

\subsection{Critical neighborhood}

We seek an expression for the physical properties of the binary
system in terms of $(T,\mh,\mdh)$ near the critical locus
$(\Tc,0,\mdh)$. For this purpose, we first solve for $\ldh$ in terms
of $T$, $\mh$ and $\mdh$: this can be done \emph{implicitly} in the
general case by invoking \eqref{cslt}, and \emph{explicitly} in the
vicinity of a critical point by implementing a perturbation scheme
at fixed $\mdh$. To this end, consider the critical point at
$\Tc(\mdh)$ and $\mh_c \pe 0$ and its vicinity defined by the two
small parameters, $t \! \varpropto (T-\Tc)$, as introduced in
\eqref{deft} and $\mh$. By construction, $u$ and $\ldh$ are small
parameters near criticality, so that the integral involved in
$\calG(\vla)$ can be computed as usual in spherical models, see
\cite{mef05,joyc72,barb&mef91}. However, a significant new feature
is that the integral is now a function of \emph{two} vanishing
parameters, $u$ and $\ldh$. The appropriate extension of the
standard critical expansion \cite{barb&mef91} yields
\begin{multline}
\label{dvlG}
    \calG (\vla) = \gz \jz^{-2} \left[ 1 - p (1 - p^\dagger \ldh) \uusg + q_0 u + \gu
    \ldh/\jz  + \gd \ldhd/\jz^2  \right. 
    \\
    + \left. \cO \left(\uusg \ldhd, \ldht \right) + o(u) \right] \vir
 \end{multline}
with coefficients $p$, $p^\dagger$, $q_0$, $g_1$, and $g_2$ which in
general still depend on $\theta$, and with the critical exponent
\begin{equation}\label{defgam}
    \gamma = \max \left[ 2/(d-2) ; 1 \right] \pt
\end{equation}
The integrals $\calLs(\vla)$ are less singular and one finds
\begin{equation}
    \label{dvlLs}
    \calLs (\vla) = \gz \jz^{-1} \left[\lsz + \lsu \ldh / \jz + \lsd \ldhd /\jz^2+ \cO
    (u,\ldht) \right] \pt
\end{equation}
We note  that $\gu$ and $\lsu$ vanish in the symmetric case when
$\dJd \pe 0$, while $\gu$, $\gd$, $p^\dagger$, $\lsz$, $\lsu$ and
$\lsd$ are all of order $\Delta J_{\sisi}/\jz$.

Using these expansions one can explicitly expand the terms in the
spherical constraints \eqref{cslt} about their values at
criticality, which then provides the required relation between
$\uusg$, $\ldh$ and $t$ and $\mh$. To proceed further, we aim to
choose the mixing parameters and, explicitly, $\thm$, to ensure that
the resulting expansion for $\uusg$ begins at orders $t$ and
$\mh^2$, as in \eqref{eqgenms}, rather than with $\mh$ as
\eqref{cslt} naively implies. This can be done by imposing the
condition
\begin{equation}
    \label{ctm}
    \thm^4 =  \frac{1-g_1(\theta)-\theta l_{2,1}(\theta)}{\theta
    [\theta  + \theta \gu(\theta) + l_{1,1}(\theta)]} .
\end{equation}

To study this, consider first the \textbf{symmetric case} when $\gu
\pe \lsu \pe 0$; the condition then  reduces simply to $\thm^2 \pe
1/\theta$, which, in combination with \eqref{md2th}, leads to the
equation
\begin{equation}
\label{mdths}
    (1 + \mdhd \lsz ) (1-\theta^2) = 0 \pt
\end{equation}
The conditions  \eqref{hjud} enforce $\dJtss \geq 0$ in symmetric
systems which leads to $l_{1,0}\pe l_{2,0} > 0$. Consequently, the
only positive solution of this equation is $\theta \pe 1$,
independently of $\mdh$. Thus we obtain $\theta \pe \thm \pe 1$.
Finally, we discover, as naturally expected, that symmetric
criticality is confined to the manifold $\mb_c \pe 0$ or
$(\rho_1+\rho_2)_c \pe \fud \volz^{-1}$, while the critical locus is
given explicitly by the simple parabolic form
\begin{equation}\label{Tcsym}
    \frac{\Tc^{\sym} (\rho_1,\rho_2)}{T^{\sym}_{c,\max}} = 1-\mdd
    = 1 - \volz^2 (\rho_1-\rho_2)^2 ,
\end{equation}
where $\kb T^{\sym}_{c,\max} / \jz \pe 1 / \gz(1)[ 1 + \lsz(1)]$.

In the general, \textbf{nonsymmetric case} the condition \eqref{ctm}
is less tractable and might even lead, one could suspect, to
multiple solutions. To keep the analysis at the simplest level, we
note that the coefficients $g_1(\theta)$ and $l_{\sigma,1}(\theta)$
are actually of order $\Jt_{\uu}(\vz)/\jz$ and $\Jt_{\dd}(\vz)/\jz$.
For the present work, we will, thus, restrict attention to systems
in which the (1,\,1) and (2,\,2) interactions are sufficiently small
relative to the (1,\,2) attractions (which, then, predominantly
drive phase separation and yield criticality). In these
circumstances the right hand side of \eqref{ctm} remains positive,
ensuring solutions for real $\thm$ and $\theta$: we then select a
positive root for $\thm$.

Supposing then, that the condition \eqref{ctm} is satisfied, the
spherical constraint in the general case has the expansion
\begin{equation}\label{u2tm}
    p \uusg \left[1+\cO \left( \uumusg,t,\mh \right) \right]=
       \ct  t + \cm \mh^2 + \cO (t\mh,\mh^3) \vir
\end{equation}
where, with the coefficients $p$, $l_{1,0}$, $\cdots$ defined via
the expansions \eqref{dvlG} and \eqref{dvlLs}, we find
\begin{equation}\label{}
    \ct =  1 +
    \frac{\thm^4 l_{1,0}+l_{2,0}}{\theta\thm^4+1/\theta}
\end{equation}
and, with
\begin{equation}
c_0 \pe \jz / g_0 \kb \Tc \quad \mbox{and} \quad w(\theta)\pe
1-g_1(\theta)-\theta l_{2,1}(\theta) ,
\end{equation}
\begin{multline}
    \cm =  \frac{2 c_0}{\theta \thm^2+1/\theta\thm^2} \\
    +  \frac{2 \theta^2 \thm^4 c_0^2}{w^2(\theta)} \mdhd
    \left[1 + 2g_1 \frac{\theta^2 \thm^4-1}{\theta^2\thm^4 +1}+2 \gd + 2
    \frac{\thm^4 l_{1,2} + l_{2,2}}{\theta\thm^4+1/\theta} \right]
    . \hspace{2cm}
\end{multline}
These expressions are derived only for $\gamma \! > \! 1$, but the
general expansion \eqref{u2tm} remains valid when $\gamma \pe 1$
with, however, different coefficients. For $\dJtss$ small enough,
$\ct$ and $\cm$ are positive (which we suppose from here on). Thus
the structure of \eqref{u2tm} leads to the usual form of the
critical singularity in the spherical model.

The second spherical field $\ldh$, which, recalling \eqref{lmlp}
enters into $u$, is given by
\begin{equation}\label{ld2tm}
    \ldh/\jz = 2 \theta \thm^2 \mh \mdh c_0 / w(\theta) + \ct' t +
    \cm' \mh^2 + \cO (t^2, \mh t, \mh^3) \vir
\end{equation}
where, in similar fashion, the coefficients are found to be
\begin{subequations}
\begin{align}\label{}
    \ct' = &   [\theta l_{2,0} - l_{1,0}/\theta]/w(\theta)
    (1+1/\theta^2 \thm^4) \vir \\
    \cm' = & \frac{\theta \thm^2 c_0}{w(\theta) (1+1/\theta^2\thm^4)}
     \nonumber \\  & \mbox{$\hspace{22mm}$}
       \times \left[  1-1/\theta^2\thm^4 + 4 \theta \thm^2 c_0 (\theta l_{2,2} -
      l_{1,2}/\theta-2 g_1)\mdhd / w^2(\theta)  \right] .
\end{align}
\end{subequations}
It should be noted that in the symmetric case, where $\theta \pe
\thm \pe 1$, both these coefficients, $\ct'$ and $\cm'$, vanish. At
this stage, having obtained expansions for $u$ and $\ldh$--- and
thus for $\lambda_1$ and $\lambda_2$\,--- as functions of $T$,
$\mh$, and $\mdh$, we are in a position to derive all the physical
properties of the system in terms of the fluid variables $T$,
$\mu_1$, $\mu_2$, and $\rho_1$ and $\rho_2$.


\subsection{Equation of state}

To calculate the equation of state, we need to rewrite the relation
\eqref{ht2mt} for the field $\hh$ using the expansions of $u$ and
$\ldh$ in terms of $t$ and $\mh$, at fixed $\mdh$, together with the
expansion
\begin{equation}\label{devlhmjz}
    \lh - \jz = (u + \ldhd)/2\jz + \cO (\ldhq, u \ldhd, u^2) ,
\end{equation}
which follows from \eqref{defu}. At this point, we resolve the
freedom to choose $\thh$ in \eqref{h2hs} by requiring that the
resulting expression for $\hh$ is minimally singular. We achieve
this by canceling the $\cO (u)$ term introduced by the factor $\mdh
(\lh - \jz)$ on the right hand side of \eqref{ht2mt}, by choosing
\begin{equation}\label{thh2th}
    \thh = \theta \thm \vir
\end{equation}
so that $\omega_-$ in \eqref{defopm} vanishes identically. With this
choice the equation of state can be written
\begin{multline}\label{ede}
    \hh - \left[\hh_c + \jz c_h t + \Ndema \mh  + \cO (\mh t, t^2) \right] =   \\
    \jz^{-1} p^{-\gamma} \mh \bigl\{ \ct t + \cm \mh^2   + \cO (\mh^3, t \mh)
    \bigr\}^\gamma 
    \left[1 + \cO(t,\mh,\uumusg )\right] \vir
\end{multline}
provided the expression in braces, which derives from $u$, remains
nonnegative; otherwise, this expression must be replaced by zero.
Recall indeed, that $u$ must be nonnegative for the free energy to
be well defined.

On the left hand of \eqref{ede} the coefficient
\begin{equation}\label{htc}
    \hh_c (\mdh) = \jz \mdh (1-\theta^2 \thm^4)/\theta \thm^2 \vir
\end{equation}
serves to specify the critical fields, $h_{1,c}$ and $h_{2,c}$ [via
\eqref{h2hs} and \eqref{devlhmjz}] and thence the critical chemical
potentials $\mu_{1,c}$ and $\mu_{2,c}$. Note that $\hh_c$ vanishes
with $\mdh$ so that in a symmetric system, where $\theta\pe\thm \pe
\theta_h \pe 1$, criticality occurs, as natural, when $\hb_c \pe 0$.

The linear term in $t$, with mixing coefficient
\begin{equation}
c_h (\mdh) = 2 \mdh \ct' ,
\end{equation}
similarly determines the near-critical $T$-dependence of the
chemical potentials, $\mu_{\Sigma,1} (T;\mdh)$ and $\mu_{\Sigma,2}
(T;\mdh)$, on the phase boundary,  a feature to be anticipated in
binary fluid mixtures.

Finally, note the linear term in $\mh$ on the left hand side of the
equation of state \eqref{ede}: this is quite unanticipated from the
perspective of previously studied spherical models, at least to the
authors' knowledge. The corresponding coefficient, which for reasons
to be explained below, we call the ``demagnetization factor,'' is
given by
\begin{multline}
\label{defNdema}
    \Ndema (\mh,\mdh) /\jz = -\frac{(1-\theta \thm^2)^2}{\theta \thm^2}+
    4 c_0 \frac{\theta \thm^2}{w(\theta)} \mdhd  
     \\ + 2 \cm' \mh \mdh + \cO
    (\mh^2) , \hspace{5cm}
 \end{multline}
in which further powers of $\mh$ should be noticed. In a symmetric
model, this factor simplifies to the fairly explicit expression
\begin{equation}
    \label{nsym}
    \Ndema  (\mb, \md) = 4 \jz (1+l_{\sigma,0})\frac{\mdd}{1-\mdd}
    \left[1 + \frac{2 \mb^2}{\bigl(1-\mdd\bigr)^2}
    + \cO (\mb^4) \right] \pt
\end{equation}

Finally, as regards the second external field $\hdh$, or chemical
potential, near criticality we have
\begin{multline}\label{hd2mt}
    \hdh - \left[\hdh_c + \Ndema' \mh + \cO (t^2, \mh t) \right] = \\
    \jz^{-1} p^{-\gamma} \mdh \biggl[\ct t + \cm \mh^2+ \cO(t\mh,\mh^3)\biggr]^\gamma 
    \left[1+\cO(t,\mh,\uumusg)\right] \vir
\end{multline}
where the right hand side has the same form and is subject to the
same conditions as in \eqref{ede} while the critical point value is
determined by
\begin{equation}\label{htdc}
    \hdh_c (\mdh) =  \jz \mdh (1+\theta \thm^2)^2/\theta \thm^2 ,
\end{equation}
and the modified demagnetization factor is
\begin{multline}
    \Ndema' (\mh,\mdh) =  \jz \frac{\theta^2 \thm^4-1}{\theta
    \thm^2} 
    \\ + 4 \jz c_0 \frac{\theta \thm^2}{w(\theta)} \mh \mdh \left( 1 +
    c_0 \frac{\theta \thm^2}{w(\theta)} \mdh   \right)
    + \cO(\mh^2). \hspace{3cm}
\end{multline}
For symmetric models, these three relations reduce simply to
\begin{equation}
\label{hdsym}
\hd = 4 \jz \md + \cO(\mb,t^2) .
\end{equation}

\subsection{Phase diagram and critical behavior}

\begin{figure}
  \includegraphics[width=8cm]{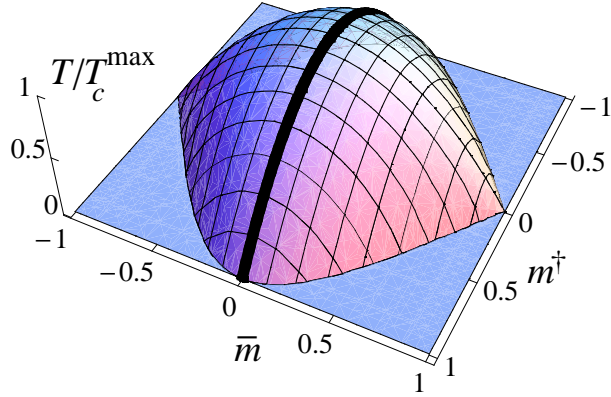}\\
  \caption{(Color online) Phase boundary in terms of the magnetization $\mb\pe \fud(m_1+m_2)$ and
  $\md \pe \fud (m_1-m_2)$ in the symmetric case.
  Criticality occurs on the bold line.
  The surface represents the limit of the single-phase region.
  Below the surface the parameter $u$ sticks at zero.}
  \label{Tmmd}
\end{figure}

Our result \eqref{ede} and subsequent relations
\eqref{htc}-\eqref{defNdema}, describe the equation of state,
i.e.~the relations between the densities, $m_1$, $m_2$, or $\rho_1$
and $\rho_2$, and the fields, $h_1$, and $h_2$, or chemical
potentials, $\mu_1$ and $\mu_2$, at temperature $T$ close to
criticality. To reveal specific, characteristic features we
consider, first, the \emph{phase boundary} in terms of the sum and
difference densities $\mh$ and $\mdh$. It follows from \eqref{ede}
that the phase boundary below and up to $\Tc (\mdh)$ is determined
by the relation $u(t,\mh,\mdh)\pe0$. Figure 2 depicts the boundary
in the space $(T,\mb,\md)$ for a symmetric system, for which the
critical locus was already derived in \eqref{Tcsym}. Evidently, at
fixed $\md$ and for $T<\Tc(\md)$ there is a composition gap $\Delta
\bar{\rho} (T) \pe \bar{\rho}_\alpha-\bar{\rho}_\beta$, where
$\alpha$ and $\beta$ label the two phases. This vanishes as
$T\rightarrow \Tc(\md)$ and from the magnetic perspective is most
readily expressed in terms of the spontaneous magnetization which is
described by
\begin{equation}
\mh_0 (T) \approx B |t|^\beta \quad \mbox{with $\beta \pe \fud$} ,
\end{equation}
where $B \pe (c_t/c_m)^{1/2}$. In fact the critical exponent $\beta
\pe \fud$ represents the standard ``universal'' spherical model
result!

As regards the other critical exponents of the general binary fluid
model, our choice of the mixing parameters $\theta$, $\thm$ and
$\theta_h$ at fixed $\mdh$ ensures that they basically match those
of the corresponding single-component spherical models. This, of
course, is in agreement with general considerations of the
thermodynamics of multi-component fluids \cite{grif&whee70}. Thus
regarding the density correlation functions, the decomposition
\eqref{natdec} shows that the dominant behavior of the density
structure function near criticality is given by
\begin{equation}\label{SNNc}
    \St_{\NN} \kTr \approx 1/(u + k^2 R_c^2 + \ldots) \vir
\end{equation}
where $R_c$ is a nonzero range parameter, while on the critical
isochore, $\mh\pe \mh_c \pe 0$, we have $u\sim t^\gamma$ as follows
from \eqref{u2tm}. Hence, we find $\St_{\NN} (\vk \pe \vz; T,
\vro_c) \! \sim \! 1/t^\gamma$ which is consistent with the
definition of the critical exponent $\gamma$ via, say, light
scattering experiments. At criticality, this result also implies
$\St_{\NN} (\vk; \Tc, \roc) \! \sim \! 1/k^2$, which confirms the
classical value $\eta \pe 0$ of the critical point decay exponent.

Moreover, in leading order close to but above criticality one can
establish the scaling form
\begin{equation}
S_{\NN} (\vk;T,\rho_c) \approx t^{-\gamma} X_{\NN} [k \xi_{\N} (T)]
,
\end{equation}
where the density correlation length is $\xi_{\N} (T)\! \approx \!
\xi_0 /t^\nu$ with critical exponent $\nu \pe \fud \gamma$ in accord
with the general scaling relation $\gamma \pe (2-\eta)\nu$. The
scaling function $X_{\NN}(x)$ has the standard Ornstein-Zernike form
$\varpropto \! 1/(1+x^2)$.

However, the equation of state in terms of $T$, $\mh$, and the field
$\hh$ requires further examination. Thus, while the standard
spherical model singularities embodied in \eqref{eqgenms} are
evident, new features arise from the ``demagnetization'' term
$\Ndema \mh$ on the left hand side of \eqref{ede}. To understand
their significance, consider symmetric systems (i.e., with
$\Jt_{\uu} \pe \Jt_{\dd}$), where \eqref{ede} reduces to
\begin{equation}
\label{edesym}
\hb \approx \Ndema \mb + \jz p^{-\gamma} (c_t t + c_m
\mb^2)^\gamma ,
\end{equation}
while the demagnetization factor $\Ndema(\mb,\md)$ is then given by
\eqref{nsym} and vanishes only if $\md \pe 0$. More generally,
however, we note that $\Ndema(\mh\pe 0,\mdh)$ vanishes in asymmetric
systems, \emph{only} on the special two loci $\mdh\pe\pm \mdh_s$
where one finds
\begin{equation}
    \mdhd_s = (1-\theta \thm^2)^2
    w(\theta)/ 4 \theta^2 \thm^4 c_0 .
\end{equation}

\begin{figure}
 \includegraphics[width=8cm]{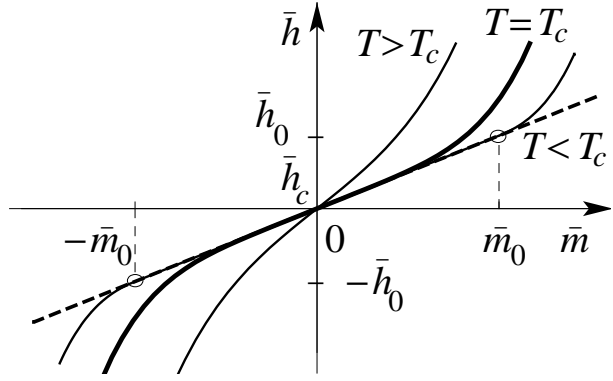}\\
  \caption{Schematic depiction of the equation of state at fixed $\mdh$ ($\neq 0$)
  in a symmetric system for temperatures above, at, and below
  criticality. In the latter case, the equation of state reduces to the linear
  demagnetization form $\hb \pe \Ndema \mb$ (dashed line)
  for $-\mb_0 \leq \mb < \mb_0$ (where $\mb_0 \sim t^\beta$) and $\hb$
  in the interval $(-\hb_0,\hb_0)$.}
  \label{figh2m}
\end{figure}

It now follows from \eqref{edesym} that the inverse thermodynamic
susceptibility or partial compressibility, $1/\chi \equiv (\partial
\hb/\partial \mb)_t$, does \emph{not} vanish at the critical point
$t\pe 0$, $\mb_c\pe 0$: see Figure \ref{figh2m}. Consequently, the
susceptibility does not diverge near criticality in the general
case! The culprit is clearly the demagnetization effect, i.e., the
term $\Ndema \mb$ which arises both from the compositional asymmetry
(when $\mdh \pneq 0$) reflecting the multicomponent nature of the
binary fluid, and, from the underlying vectorial character of the
order parameter: recall that spherical models correspond to the
$n\rightarrow \infty$ limit of systems of vector-valued spins.
Indeed, as we demonstrate in the next section, the origin of the
demagnetization effect in spherical models can be understood
directly in terms of vector spin models.

The nondivergence of the susceptibility/compressibility $\chi$ means
that standard isothermal plots of the chemical potential (or,
similarly, the pressure) vs.~density or of magnetic field
vs.~magnetization near criticality take the form illustrated in
Fig.~\ref{figh2m} with, in general, a nonzero slope \emph{at}
$T\pe\Tc$ fixed by the value, $\Ndema_c$, of the demagnetization
coefficient. Note, in particular, that below $\Tc$ what would in a
standard fluid system be a constant isotherm of zero slope through
the two-phase region\,--- i.e., the interval set by the spontaneous
magnetization, $m_0(T)$\,--- is now replaced by a straight line with
the same fixed slope $\Ndema_c$ (at least close to $\Tc$).

It is this fact that leads us to call this anomalous behavior,
certainly unphysical in a fluid model, a ``demagnetization effect''.
To be more specific, in a real magnetic system with long-range
dipole-dipole interactions, one must distinguish between the
externally applied magnetic field $H_{\sext}$, analogous here to
$\hb$ (or $\hh$), and the \emph{internal field}, $H_{\sint}$, which
is what is ``seen'' by individual atomic and molecular spins. The
relation between these may be written
\begin{equation}
H_{\sint} = H_{\sext} - N M , \label{eqdema}
\end{equation}
where $M$, analogous here to $\mb$ (or $\mh$) is the magnetization
while $N$ is the demagnetization coefficient
\cite{Star41,LandLifs60,Matt65}. More generally the fields
$\mathbf{H}_{\sint}$ and $\mathbf{H}_{\sext}$ are real-space
vectors, as is $\mathbf{M}$, and the relation \eqref{eqdema} can be
used only when the system is in the form of an ellipsoid and
directions parallel to the major axes are considered. (In the case
of a sphere one has $N \pe 4 \pi /3$
\cite{Star41,LandLifs60,Matt65}.)

One might, in light of these considerations, ask if one should not,
similarly, be able to introduce an ``effective internal field'',
$\hh_{\sint} \approx \hh_{\sext} - \Ndema \mh$, that would play a
natural thermodynamic role. However, on the one hand, given the
implicit variation of $\Ndema(\mh,\mdh)$, this seems unlikely to be
related to the basic thermodynamic parameters, $T$, $\mu_\sigma$,
and $\rho_\sigma$, sufficiently directly to be of real value, and,
on the other hand, the higher order terms in \eqref{ede} indicate
that the linear slope shown in Fig.~\ref{figh2m} for the
``two-phase'' region, will become \emph{nonlinear} outside the
critical region.

Finally, as a further caution, another unphysical feature of the
present multicomponent fluid models must be noted. Indeed, it enters
even in single-component spherical models
\cite{mef05,jn&mefrmp,mef92}! Specifically, whenever $\gamma \! > \!
1$ and $\Ndema_c \pe 0$ so that the susceptibility $\chi$ diverges
to $\infty$ on approach to $\Tc$ along the critical isochore, it
\emph{also} diverges when the phase boundary is approached below
$\Tc$. Even for $\Ndema_c \pneq 0$, a corresponding anomalous
feature arises and is embodied in Fig.~\ref{figh2m} where the
\emph{slope} of the isotherm below $\Tc$ remains \emph{continuous
through} the phase boundaries (marked by open circles); but in
realistic fluid models there would be breaks in the slope!




\section{Vector spin model analysis}
\label{model}

To understand the origin of the demagnetization-like effects that
enter the present multi-component spherical models, it is helpful to
recognize that spherical models correspond precisely to the $n \!
\rightarrow \! \infty$ limit of appropriate systems of $n$-component
spins \cite{stan68}. The vectorial character of the order parameter
is thus a trademark of the model \cite{mef05,jn&mefrmp}, coupled
here to asymmetry and multicomponent features. The `secret' of the
demagnetization effect appearing first in \eqref{ede} can then be
understood by regarding our binary-fluid spherical models as
magnetic models with two classes of spins on separate sublattices,
just as in \eqref{Xi}-\eqref{deftfJst}, but now as fixed length
\emph{vector spins}, $\mathbf{s}_\sigma (\vRis)$ ($\sigma\pe 1, 2$),
rather than scalar Ising spins as originally contemplated.

To obtain insight into the behavior of the model below $\Tc$, we may
use a simple mean-field approach by representing the overall
sublattice magnetizations by two mean values, $\vm_1$ and $\vm_2$.
The lengths of these magnetization vectors should ideally be taken
as $m_\sigma^0 (T)$, the spontaneous magnetizations (at fixed $T \!
< \! \Tc$); but it suffices here to consider the symmetric situation
and so accept equal fixed lengths $|\vm_1|\pe |\vm_2| \pe 1$.

\begin{figure}
  \includegraphics[width=7cm]{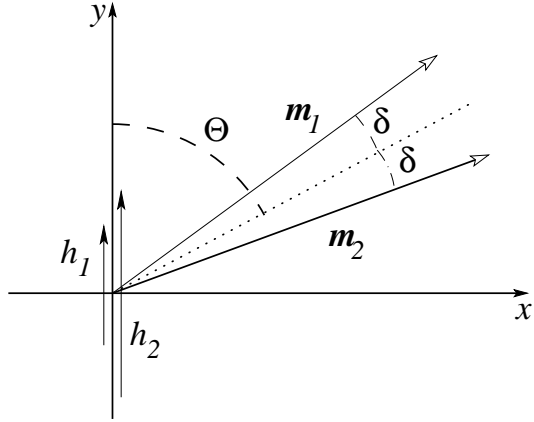}\\
  \caption{Vectorial representation of the two-species spherical model within
  a mean-field picture in which $\vm_1$ and $\vm_2$ are the two sublattice
  magnetization vectors that can rotate with respect to the direction,
  the $y$ axis, set by the parallel external fields, $\vh_1$ and $\vh_2$.}\label{figxy}
\end{figure}

Then, in contrast to most realistic magnetic systems, it is
imperative to allow for the imposition of two distinct external
magnetic fields $\vh_1$ and $\vh_2$, corresponding, as fundamental
for fluid models, to two distinct chemical potentials $\mu_1$ and
$\mu_2$. However, for the ``chemical interpretation'', we must take
$\vh_1$ parallel to $\vh_2$ and may identify the preferred direction
as the $y$ axis: see Fig.~\ref{figxy}. The components $(\vm_1)_y$
and $(\vm_2)_y$ then correspond to the densities $m_1$ and $m_2$ in
our previous analysis, while
\begin{equation} \hb \pe \fud
(h_1+h_2) \quad \mbox{with} \quad h_\sigma \pe (\vh_\sigma)_y
\end{equation}
describes the external field/chemical potential. On the other hand,
$\hd \pe \fud (h_1-h_2)$ characterizes the compositional
\emph{asymmetry} of the system (even when the two species, here the
magnetic sublattices, are symmetrically related): that asymmetry is
at the heart of the matter. As regards the vector-spin
dimensionality, however, it suffices to allow for only one more
dimension and so, regard the $\vm_\sigma$ as XY or O$(2)$ order
parameters.

Finally, beyond the symmetric \emph{intra}sublattice ferromagnetic
couplings (that lead to the spontaneous magnetizations), we allow
for the \emph{inter}sublattice interactions by a coupling constant
$j\! > \! 0$ (analogous to $\jz$ above). Thus we take the essential
part of the mean-field free energy to be
\begin{equation}\label{}
    \cF (\hb; \hd; \vm_1, \vm_2) = - \vh_1 \mathbf{\cdot} \vm_1 - \vh_2 \mathbf{\cdot} \vm_2
    - j \vm_1 \mathbf{\cdot} \vm_2 .
\end{equation}
Here the external field $\hb$ is the control variable while $\hd$ is
fixed and, as usual, $\cF$ is to be minimized with respect to
$\vm_1$ and $\vm_2$.

Let us, as in Fig.~\ref{figxy}, introduce the mean tilt angle
$\Theta$ between the $y$-axis and $\bar{\vm} \pe \fud (\vm_1+\vm_2)$
and the splitting or separation angle $\delta$, between the
$\vm_\sigma$ and the $\bar{\vm}$. For simplicity we suppose that
$\delta$ is small (which requires $|\hd|\ll j$); then minimization
on $\delta$ yields
\begin{equation}\label{Eminth}
    \cF_{\min} (\Theta) = - j - 2 \hb \cos \Theta - \frac{2 \hdd
    \sin^2 \Theta}{2 j + \hb \cos \Theta} + \cO (\hdd/j^2) .
\end{equation}

Consider, first, the fully symmetric case in which $\hd \pe 0$ (and
$\delta_{\min} \pe 0$). Minimizing this expression on $\Theta$ then
gives $\Theta_{\min} \pe 0$ for $\hb \pe 0$ but $\Theta_{\min} \pe
\pi$ for $\hb \! < \! 0$. This evidently corresponds to the usual
ferromagnetic situation in which (\emph{neglecting} dipolar
interactions and demagnetization effects) the magnetization
$\bar{\vm}$ switches abruptly from $\mb \equiv (\bar{\vm}_\sigma)_y
\pe -1$ to $+1$ as the field $\hb$ passes through zero.

On the other hand when $\hd  \pneq 0$ the minimizing value of
$\Theta$ assumes a nontrivial value for $\hb$ between the limits
$\pm \hb_0$ given by
\begin{equation}\label{hc}
    \hb_0 = j \left( \sqrt{1+\hdd/j^2}-1\right) \approx \fud \hdd / j
    ,
\end{equation}
up to corrections of relative order $(\hd/j)^2$. As a consequence,
the magnetization $\mb(\hb)$ no longer jumps discontinuously at $\hb
\pe 0$ from $\mb \pe -1$ to $\mb \pe +1$ but rather varies
continuously and almost linearly over the interval $-\hb_0 \! \leq
\! \hb \! < \! \hb_0$ according to
\begin{equation}
\label{mbnl}
 \mb (\hb) \simeq \frac{\hb}{\hb_0} \left[ 1-\mbox{$\frac{3}{8}$} (\hd/j)^2\right]
 / \left[1-\mbox{$\frac{3}{8}$}(\hd/j)^2 (\hb/\hb_0)^2
\right] .
\end{equation}
More explicitly to leading order in $(\hd/j)^2$ one finds for
$|\hb|\! < \! \hb_0$,
\begin{equation}\label{x2hhd}
    \mb =  \cos \Them = \frac{2 j}{\hb} \left[
    \sqrt{\frac{1-\hb^2/4j^2}{1-2\hb^2/\hdd}} - 1 \right] \vir
\end{equation}
while for $|\hb|\! > \! \hb_0$ one has $\cos \Theta \pe \textrm{sgn}
\{\hb \}$.

In words, for an external field $\hb$ not too large compared to the
square of the asymmetric field $\hd$, the spins cant themselves in a
direction $\Them \neq 0$, with, indeed, $\Them \pe \pi /2$ when $\hb
\pe 0$! This minimizing behavior of vector spins is clearly the
origin of the seeming demagnetization effect and explains our result
for the spherical model. Indeed, near the origin, for $\hb/\hb_0
\rightarrow 0$, we find $\cos \Them \approx 2 \hb j / \hdd$ which
leads to a nondivergent susceptibility of magnitude
\begin{equation}\label{}
    \chi \pe \left(
\partial \mb / \partial \hb \right)_{T, \hb = 0} = 2  j / \hdd .
\end{equation}
Thus, as the spherical model itself, the asymmetry of the spin
model, coupled to the vectorial character of the order parameter,
leads to a non-vanishing inverse susceptibility near criticality in
the general case. Note also that, as in the spherical model, the
divergence of $\chi$ re-emerges in the symmetric case when $\hd \pe
0$. Finally, the nonlinear terms in $\hb$ implied by \eqref{mbnl}
show that one cannot hope to find a simple demagnetization
description as in \cite{Star41,LandLifs60,Matt65}.

\section{Conclusions}
\label{conc}

We have introduced multicomponent generalizations of the standard
spherical model that embody lattice-gas hard cores for many-species
fluids by using interlaced sublattices. Taking into account a
spherical constraint for each distinct species, we have obtained
exact expressions for the free energy and pair correlation functions
in the general case, in terms of the basic Fourier space interaction
matrix. We have then focused on binary fluids where the
diagonalization of $2 \! \times \! 2$ matrices leads to relatively
simple results for the physical properties of the system. We find
that density and (for ionic fluids) charge correlation functions can
be decomposed naturally in terms of two eigenmodes. This
formulation, which could well have broader validity, has dramatic
consequence for charged fluids as we expound elsewhere
\cite{mef05,jn&mefrmp,jn&mef04PRL,jn&mef04JPA,jn&mefsfp}.

The present article considers fluids where, in addition to hard
cores, only short-range overall attractive interactions are present.
We show that with an appropriate choice of variables (in the form of
linear combinations of the mean magnetizations/densities or external
fields/chemical potentials for the two species), the usual critical
properties of single-component spherical models can be uncovered in
accord with general thermodynamic arguments for multicomponent
solutions. Specifically, as the relative composition of the system
varies, criticality is realized on a well defined locus in the full
phase diagram.

However, an unexpected and intrinsically unphysical
``pseudodemagnetization phenomenon'' arises that, except on certain
submanifolds, prevents the usual divergence of the thermodynamic
susceptibilities/compressibilities at criticality. This feature,
undesirable for model fluids, is found to be a consequence of an
interplay between species and compositional asymmetry and the
multidimensional characteristics represented by the hidden vectorial
character of the order parameter in spherical models. The behavior
can, indeed, be understood via a simple mean-field description of a
corresponding XY spin model. Despite this artificial aspect,
requiring caution in interpreting results, further aspects of the
multicomponent spherical models seem worth pursuing: on the one
hand, ternary fluid models with, say, positive and negative ions in
solutions of neutral molecules, could be instructive; on the other
hand, magnetic systems with different types of ions could reveal
interesting behavior.

\begin{acknowledgments}
In the period 2002-2004, this work was supported through the
National Science Fondation under Grants No 99-81772 and 03-01101.
J.-N. A. appreciates both support and hospitality from the Institute
for Physical Sciences and Technology at the University of Maryland.
\end{acknowledgments}


\appendix
\section{Location of singularities}
\label{appA}

In this appendix, we locate the singularities of the binary
spherical model for the generic case of short-range interactions.
The singularities derive from the vanishing of $\Lm \kl$ and we
obtain sufficient conditions to ensure that they arise only when
$\vk \pe \vz$.

\subsection{Vicinity of the origin}

We first consider the small-$\vk$ behavior of $\Lm \kl$. Owing to
the second condition \eqref{hjud}, the range $\Rud$ defined in
\eqref{dljst} satisfies $\Rud^2 \! > \! 0$. Note also the relations
$\dJb (\vk) \! \approx \! \jz \Rb^2 k^2$ and $\dJd (\vk) \! \approx
\! \jz {\Rd}^2 k^2$, where $\Rb$ and $\Rd$ are characteristic
ranges. At leading order in $k$ we then have
\begin{equation}\label{dvllm}
    \frac{\Lm \kl - \Lm \zl}{\jz k^2} = \Rb^2 +
    \frac{ \jz \Rud^2  -\ld  {\Rd}^2}{\rlddjzd} + \cdots ,
\end{equation}
which will be valid in a domain $|\vk| \leq k_m \! > \! 0$. From the
third condition \eqref{hjud} we find $\Rb^2 \! > \! 0$ and if,
recalling the definition \eqref{defdJd} of $\dJd$, we accept the
further condition
\begin{equation}
\label{cRd} |{\Rd}^2| = \lim_{k \rightarrow 0} \dJd (\vk)/\jz k^2
\leq \Rb^2 ,
\end{equation}
we are assured that $\vk \pe \vz$ is indeed the minimum of $\Lm \kl$
when $k \leq k_m$.


\subsection{Remainder of the Brillouin zone}

Consider now the subdomain $\B'$ of the Brillouin zone $\B$,
consisting of all vectors $\vk$ outside the origin domain $|\vk|
\leq k_m$. In the symmetric case when $\dJd \pe 0$ (marked by
superscripts $\sym$), the second condition \eqref{hjud} leads to
$|\Jt_{\ud}(\vk)|\! < \! \jz$ so that
\begin{equation}\label{}
    \Delta \Lm^{\sym} \equiv \Lm^{\sym} \kl - \Lm^{\sym} \zl > \dJb (\vk) \pt
\end{equation}
But, according to the last member of \eqref{hjud}, there exists a
$\dLm^{\sym}$ such that
\begin{equation}\label{}
\Delta \Lm^{\sym} \kl \geq \dLm^{\sym} > 0 \vir \, \, \textrm{for
$\vk \in
 \B'$} \pt
\end{equation}
In the general case where $\dJd$ is arbitrary, we may write
\begin{equation}\label{}
    \Lm \kl - \Lm \zl = \Delta \Lm^{\sym} \kl + \delta \Lm [\dJd (\vk)] \vir
\end{equation}
where from \eqref{lpm} we define
\begin{equation}
    \delta \Lm [\dJd (\vk)]  = \sqrt{\ldd+\fuq \Jtud^2 (\vk)} 
     - \sqrt{[\ld+ \dJd(\vk)]^2 + \fuq \Jtud^2 (\vk)} \pt
\end{equation}
Then, noting that $|\partial(\delta\Lm)/ \partial \dJd (\vk)  | \leq
1$ and $\dJd(0)\pe 0$, we see that $|\delta \Lm [\dJd(\vk)] | \leq
|\dJd(\vk)|$. Hence, accepting the further condition
\begin{equation}\label{hdjd}
    \max_{\vk \in \B'} \left[ |\dJd (\vk)| \right] < \dLm^{\sym} \vir
\end{equation}
which means that the asymmetry is not too strong,  one concludes
that $\Lm \kl \! > \! \Lm \zl$ for all $\vk$ in $\B'$.


\end{document}